\documentclass{iau}
\usepackage{graphicx}

\title[IAUS293.~~Nanosatellites and exoplanet detection] 
{Do have nanosatellites a role in detecting exoplanets?}

\author[Werner W. Weiss, Vera Maria Passegger \& Jason Rowe]   
{Werner W. Weiss$^1$, Vera Maria Passegger$^1$
 \and Jason Rowe $^2$}

\affiliation{$^1$University of Vienna, Institute for Astrophysics, Tuerkenschanzstr. 17,  Vienna, 1180 Austria \\
                 $^2$NASA Ames Research Center, CA 94035-0001 Moffet Field, USA
\\email: {\tt werner.weiss@univie.ac.at}}

\pubyear{2013}
\volume{293}  
\pagerange{?--?}
\jname{Formation, detection, and characterization of \\extrasolar habitable planets}
\editors{NN, eds.}
\begin{document}

\maketitle

\begin{abstract}
In December 2012, Austria will launch its first two satellites: UniBRITE and BRITE-Austria. This is the first pair of three, forming a network called BRITE-Constellation. The other pairs being contributed by Canada and Poland. The primary goal of BRITE-Constellation is the exploration of short term intensity variations of bright stars (V$>$6 mag) for a few years. For each satellite pair, one will employ a blue filter and the other a red filter. With the discovery of the first exoplanet in 1992, more than 800 have been detected since. The high-precision photometry from the BRITE instrument will enable a transit search for exoplanets around bright stars. 

To estimate the capability of BRITE to detect planets, we include in our calculations technical constraints, such as photometric noise levels for stars accessible by BRITE, the duty cycle and duration of observations. The most important parameter is the fraction of stars harboring a planet. Our simulation is based on 2695 stars distributed over the entire sky. Kepler data indicate that at minimum 34\% of all stars are orbited by at least one of five different planetary sizes: Earth, Super-Earth, Uranus, Jupiter and Super-Jupiter. Depending on the duty cycle and duration of the observations, about six planets should be detectable in 180 days, of which about five of them being of Jupiter size. 
\end{abstract}

\firstsection 

\section{Instrumentation}
BRITE-Constellation, standing for Constellation of BRIght Target Explorers, is a project consisting of 6 nanosatellites from Austria, Canada and Poland. Such a satellite cluster is needed to optimize the duty cycle and to obtain color information (with three satellites having blue and three having red filters). 

\begin{figure}
\begin{center}
 \includegraphics[width=4.5in]{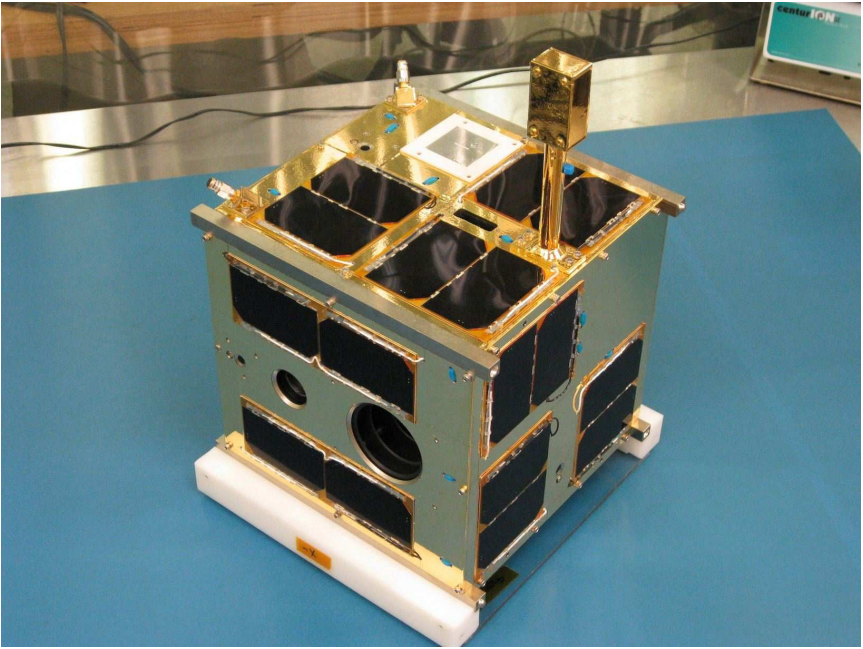} 
 \caption{UniBRITE ready for tests at the Space Flight Laboratory (SFL) of the University of Toronto.}
   \label{fig1}
\end{center}
\end{figure}

Each BRITE bus is a 20\,cm cube and has a nominal mass of 6\,kg and is based on developments of the Space Flight Laboratory (SFL) of the University of Toronto, Canada. The innovative components of this nanosatellite technology are three orthogonal reaction wheels for a three-axis high precision attitude control system and three orthogonal magnetorquer coils for momentum dumping. Attitude determination is provided by a magnetometer and six sun sensors, each of which is equipped with coarse and fine sun-sensing elements. The bus also carries a nanosatellite star tracker. This equipment will enable attitude determination to 10 arcseconds, attitude control accuracy to better than a degree, and attitude stability to within one arcminute rms. The science payload of the satellite  consists of a 5-lens camera and the interline CCD detector KAI 11002 from Kodak with 11\,Mpixel, along with a 14-bit analog to digital converter. UniBRITE, representative for the other 5 BRITE nanosatellites, ready for tests at SFL is shown in Fig.\,\ref{fig1}.

The launch of the first two satellites (the Austrian UniBRITE and BRITE-Austria) will be provided by the Indian Space Research Organization (ISRO) in Sriharikota with a current launch date in Q1/2013. Both nanosatellites will be released into a sun-synchronous polar dawn-dusk, low earth orbit 820 km above ground. The orbit period will be 101 minutes and stars in a 24$^o$ field-of-view can be observed during 15 to 40 minutes per orbit. About 2 to 20 primary target stars (brighter than V = 4\,mag) will be observable in a given field-of-view and up to about 200 more secondary target stars brighter than V = 6\,mag. 

\begin{figure}[h]
\begin{center}
 \includegraphics[width=5in,scale=1.5]{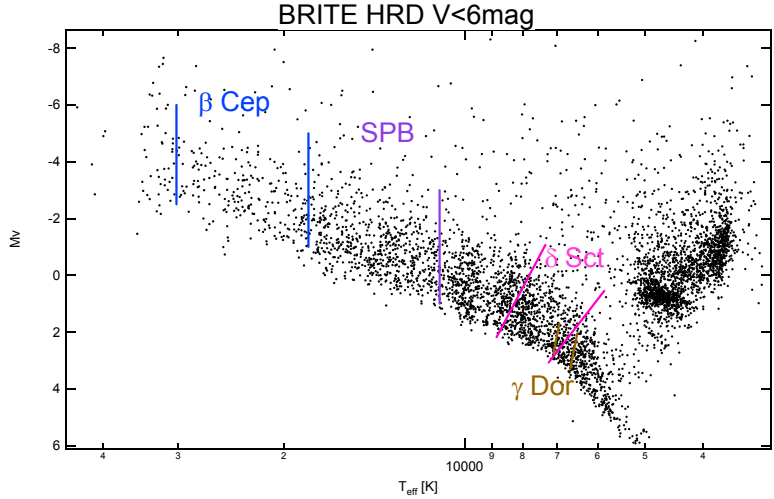} 
\vspace*{-1.3 cm}
 \caption{Hertzsprung-Russell diagram of the BRITE-sky (mag(V)$<$6)}
   \label{fig2}
\end{center}
\vspace*{1.5cm}
\end{figure}

The BRITE-Constellation mission is dedicated to a two-color high-precision photometry, which is hard to achieve from the ground for bright stars due to limiting effects of the terrestrial atmosphere. Such observations need to be done with dedicated satellites in space.  Figure\,\ref{fig2} shows the HRD of the "BRITE-sky" and illustrates the excellent coverage of the entire stellar parameter space. This HRD includes among other ''special" targets 260\,OB, 167\,CP, 75 $\beta$\,Cep, 37\,Be, 26\,EB, and 30\,$\delta$\,Scuti stars.

The primary scientific goals of BRITE are photometric measurements of oscillations, activities, temperature variations and rotation with periods between 0.02 and 100 days for single stars and eclipsing binaries. Furthermore the role of stellar winds on the stellar evolution will be analyzed and also potentually transitting exoplanets. 

\begin{figure}[h]
\begin{center}
 \includegraphics[width=5.5in]{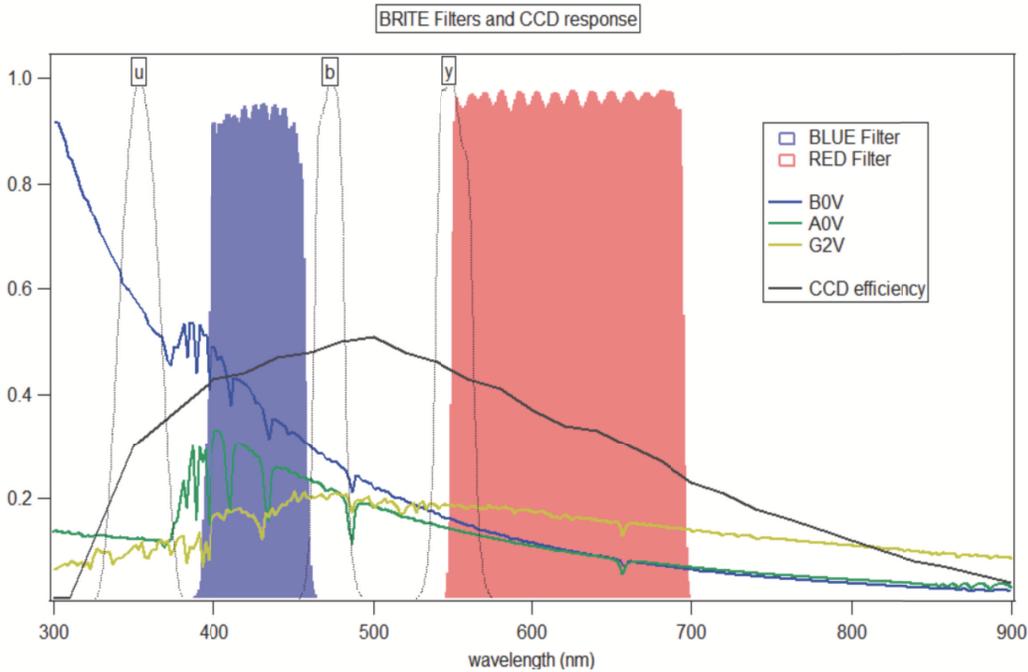} 
 \caption{Blue and red filter used for the BRITE instruments compared to the flux distribution of B-, A- and G-type stars, and to u,b and y Stroemgren filter transmission.}
   \label{fig4}
\end{center}
\end{figure}

UniBRITE will observe in the red band from 550-700 nm, BRITE-Austria in the blue band from 390-460 nm. The filters are designed such that a star with about 10\,000\,K (which is the average for the BRITE-sky) will give the same photometric signal in both filters. Furthermore, the central filter-wavelengths are separated as far as possible, considering the spectral efficiency characteristics of the CCD detector (see Fig.\,\ref{fig4}).

Stellar parameters like  temperature, luminosity, inner structure, mass, age, evolutionary status and variability  can be restricted  with lightcurves in two colors. Observations of transits in two filters are also very useful, as a transit due to an exoplanet will have similar depths in each filter. This allows for the identification of false positives, such as background eclipsing binaries, that will contaminate the survey.

\section{BRITE and exoplanet search}
To estimate the probability of BRITE to detect planets, software was developed by JR and optimized for BRITE-Constellation by VMP which estimates the number of observable planets under certain assumptions. Data parameters that must be considered in this planet search are the duty cycle, duration and transit-signal-to-noise ratio (S/N) of the  observations, and the field-of-view (FOV).  The duration of the observations is needed to estimate how many transits of a certain planet can be observed, depending on its period. The longer the data string, the more transits will be seen which increases the probability of identifying a transiting system. The most important parameter is the fraction of stars harboring a planet. Kepler data indicate that at least 34\% of all stars are orbited by at least one of five different planetary types: Earth, Super-Earth, Uranus, Jupiter and Super-Jupiter.

Equally important for detecting a transit are the duty cycle and the S/N of the data, which latter depends mainly on the brightness of the target star. The duty cycle is a measurement of how long the satellite is observing the target. The duty cycle influences the transit S/N which also depends on the planet orbit period. The higher the duty cycle, the higher the S/N of the transit. The FOV parameter simply determines the number of candidate stars which can be observed simultaneously. 

Although the primary science goal of BRITE-Constellation is not the detection of exoplanets, such a search is of increasing interest. The following analysis has been done for the BRITE-sky with the assumptions of a duty cycle of 40\%, the duration of the observation of 180 days and a cadence of 1 data point per minute. The following Table\,1 gives the number of different planet types detectable for different S/N bins. 

Please remember that the BRITE-sky contains a specific mixture of spectral types (stellar radii) depending on the line-of-sight. Kepler data may indicate a probability for exoplanets depending on spectral type and type of exoplanet (\cite[Borucki et al. 2011]{Bor}). All this information was considered when generating predicted detection rates.

\begin{figure}[b]
\begin{center}
 \includegraphics[width=5.5in]{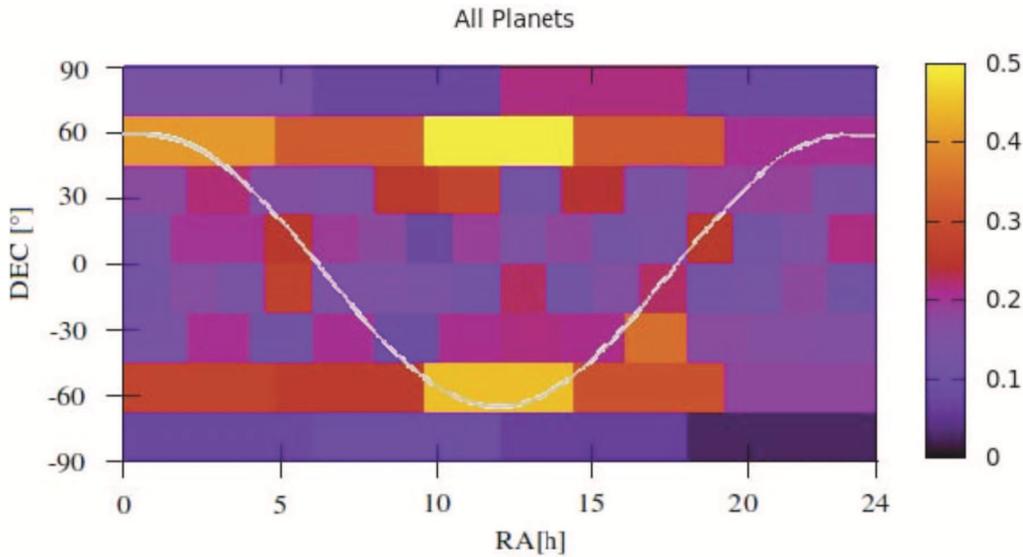} 
 \caption{Detectability of exoplanets in the BRITE-sky. The color code indicates the number of estimated detections during 180 days within the field of view of 24 degrees.}
   \label{fig3}
\end{center}
\end{figure}

Figure\,\ref{fig3} shows a map of the BRITE-sky based on a FOV of 24$^o$, color codes according to the number of planets expected in a given direction in the sky for transits with a $S/N > 5$. The galactic plane is indicated by a gray curve.

\section{Conclusion}

Table\,1 demonstrates that it is possible to detect transiting planets with BRITE. A transit-signal-to-noise ratio (S/N) of $ > 5$ should be achieved in order to be confident that the detected signal is a transit. A duty cycle of 40\% may be a bit optimistic with a duty cycle between 20-40\% being more realistic. For the duration of the observations an average of 180 days was used, which is limited by the angle between the Sun and the line-of-sight. For extended observations we expect a significant increase of straylight due to the Sun.
More accurate limits will be known after the commissioning phase. Another limiting factor for the straylight is the daylight side of the Earth which contribution is also not yet known. However, observations spanning 240 days may be possible for targets close to the ecliptic poles. 

Earth- and Uranus-sized planets may not be detectable, because of their small radius compared to their host stars resulting in a low S/N. They may be numerous in the BRITE-sky, but extremely difficult to verify in the noise. These planets might only be detectable for smaller K or M main-sequence stars. Jupiters and Super-Jupiters are much easier to detect due to their larger radii resulting in a larger S/N. Super-Jupiters have the largest radii, however, they are rare and have a large false-positive rate.  Most planets can be found along the galactic plane, as the star density is large. Also, the odds of a background blend becomes larger towards the galactic plane. This is the reason Kepler does not observe in the galactic plane. At the ecliptic northern polar region detectability increases for Jupiter and Super-Jupiter sized planets due to the larger amount of main-sequence stars in this region relative to the galactic plane. 

\begin{table}[h]
  \begin{center}
  \caption{Expected number of exoplanets. The rows show the number of expected detectable planet types for the entire BRITE-Sky with 2695 stars, assuming a duty cycle of 40\%, observations during 180 days and a cadence of 1 data point per minute. The columns refer to bins of different signal-to-noise (S/N) ratios for the transits. The last row contains the total number of expected planet detection for different S/N bins. }
  \label{tab1}
\vspace*{0.5cm}
  \begin{tabular}{|c|c|c|c|c|c|c|c|}
\hline 
{\bf S/N:} & {\bf $<1$} & {\bf $1-2$} & {\bf $2-3$} & {\bf $3-4$} & {\bf $4-5$} & {\bf $5-6$} & {\bf $>6$} \\ 
\hline
Earth             & 21.18 & 0.03 & 0.01 & 0.01 & 0.01 & 0.02 & 0.03 \\
Super-Earth  & 19.19 & 0.23 & 0.03 & 0.02 & 0.03 & 0.08 & 0.13 \\
Uranus          & 23.17 & 9.57 & 3.82 & 1.84 & 0.92 & 0.47 & 0.76 \\
Jupiter           & 0.20   & 0.56 & 0.60 & 0.85 & 0.90 & 0.74 & 4.44 \\
Super-Jupiter & 0.00  & 0.00 & 0.00 & 0.01 & 0.01 & 0.01 & 0.38 \\ 
\hline
Total             &63.75  &10.38& 4.46 & 2.73 & 1.86 & 1.32& 5.74 \\
\hline
  \end{tabular}
 \end{center}
 \end{table}

\end{document}